\documentclass[aps,preprint,preprintnumbers,prb]{revtex4}
\usepackage{amsfonts}
\usepackage{amsmath,amsbsy,amssymb,graphicx}
\let\address=\affiliation

\begin{document}

\title{
Graphene Nanoribbon and Graphene Nanodisk
}

\author{Motohiko Ezawa}

\address{Department of Physics, University of Tokyo, Hongo 7-3-1, Tokyo 113-0033, Japan}

\begin{abstract}
We study electronic properties of graphene derivatives which have closed edges.
They are finite-length graphene nanoribbons and graphene nanodisks. 
No metallic states are found in finite-length zigzag nanoribbons 
though all infinite-length zigzag nanoribbons are metallic. 
We also study hexagonal, parallelogrammic and trigonal nanodisks with zigzag or armchair edges. 
No metallic states are found in these nanodisks either except trigonal zigzag nanodisks.
It is interesting that we can design the degeneracy of the metallic states arbitrarily in trigonal zigzag nanodisks by changing the size. 
\end{abstract}

\maketitle

\textit{Introduction}: Graphene based nanostructure may be an alternative to
silicon based mesostructure in future electronic devices. Among them
graphene nanoribbons\cite{Fujita,EzawaPRB,Brey,Son,Barone,Berger} have
attracted much attention due to a rich variety of band gaps, from metals to
wide-gap semiconductors. It is intriguing that all zigzag nanoribbons have
zero-energy states\cite{Fujita,EzawaPRB} and hence they are metallic.

In order to make nanoelectronic circuits, however, nanoribbons must have
finite length. It is important to investigate the finite-length effects on
the electronic properties of nanoribbons. In this paper, we study whether
finite-length zigzag nanoribbons have zero-energy states to know if they are
metallic or not.

Graphene has a two-dimensional structure, while a graphene nanoribbon has a
one-dimensional structure. We may likewise consider a zero-dimensional
structure, that is, a graphene nanodisk. A graphene nanodisk is a
nanometer-scale disk-like material characterized by a discrete energy
spectrum. Some of nanodisks have already been manufactured by soft-landing
mass spectrometry\cite{Rader}. In this paper, we also study the electric
properties of typical nanodisks in quest of zero-energy states. A
combination of nanoribbons, nanodisks and other graphene derivatives is a
promising candidate of nanoelectronic circuits\cite{EzawaPhysica}.

\begin{figure}[t]
\begin{center}
\includegraphics[width=0.5\textwidth]{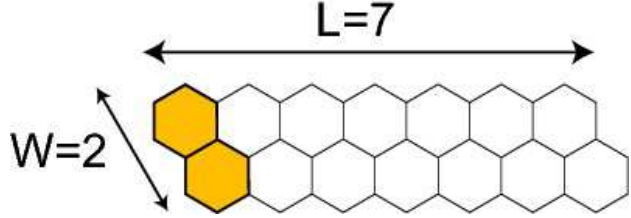}
\end{center}
\caption{(Color online) Geometric configuration of zigzag graphene
nanoribbons with width $W$ and length $L$. Here we show the example of the $%
(W,L)=(2,7)$ nanoribbon. The basic chain is $W$ connected benzene depicted
in orange (gray).}
\label{FRibbon}
\end{figure}
\textit{Finite-length graphene nanoribbons}: A classification of
infinite-length nanoribbons is given in a previous work\cite{EzawaPRB}. In
this paper we concentrate on finite-length zigzag nanoribbons. We classify
them as follows (Fig.\ref{FRibbon}). First we take a basic chain of $W$
connected carbon hexagons, as depicted in orange (dark gray). Second we
translate this chain. Repeating this translation $L$ times we construct a
nanoribbon indexed by a set of two integers $\left( W,L\right) $. In what
follows we analyze a class of finite-length nanoribbons generated in this
way. Parameters $W$ and $L$ specify the width and the length of nanoribbons,
respectively. The infinite-length nanoribbons are obtained by letting $%
L\rightarrow \infty $.

\begin{figure}[t]
\begin{center}
\includegraphics[width=0.9\textwidth]{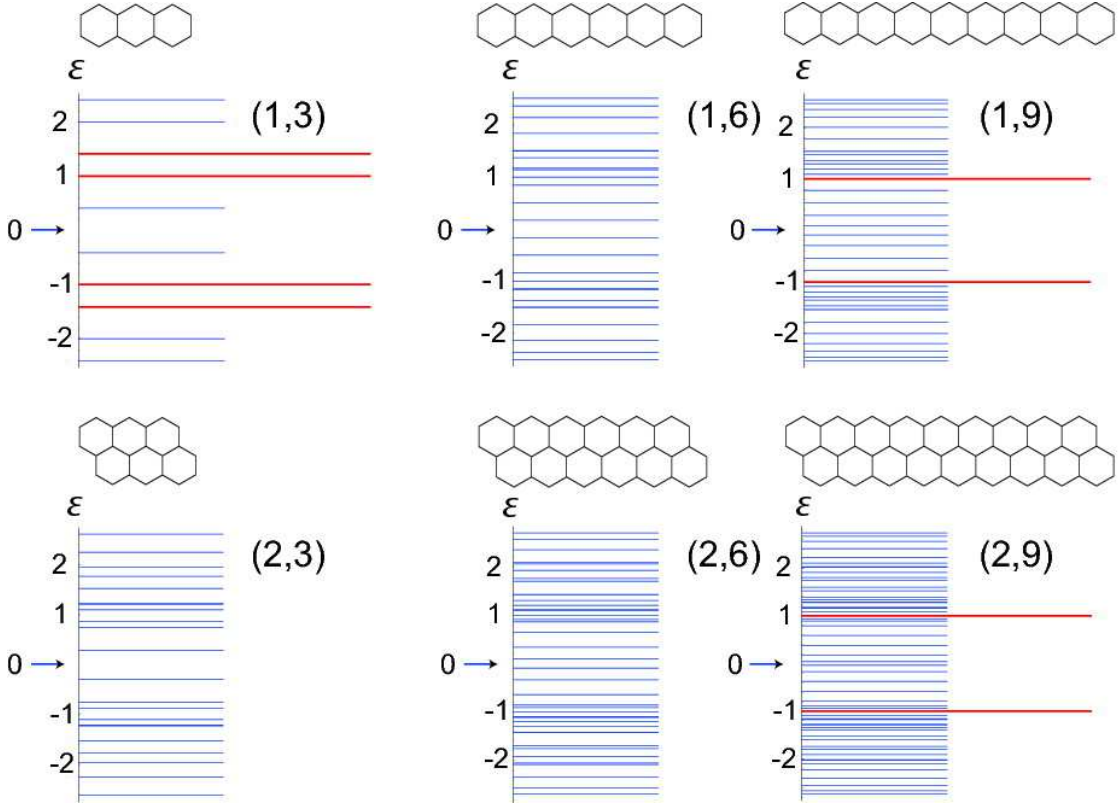}
\end{center}
\caption{(Color online) Density of states of finite-length nanoribbons. The
vertical axes is the energy $\protect\varepsilon $ \ $\ $in units of $t=3$eV
and the horizontal axes is the degeneracy. There exist no zero-energy
states. }
\label{FDOS}
\end{figure}

\textit{Energy spectrum:} We calculate the energy spectrum based on the
nearest-neighbor tight-binding model, which has been successfully applied to
the studies of carbon nanotubes and nanoribbons. The Hamiltonian is defined
by%
\begin{equation}
H=\sum_{i}\varepsilon _{i}c_{i}^{\dagger }c_{i}+\sum_{\left\langle
i,j\right\rangle }t_{ij}c_{i}^{\dagger }c_{j},  \label{HamilTB}
\end{equation}%
where $\varepsilon _{i}$ is the site energy, $t_{ij}$ is the transfer
energy, and $c_{i}^{\dagger }$ is the creation operator of the $\pi $
electron at the site $i$. The summation is taken over all nearest
neighboring sites $\left\langle i,j\right\rangle $. Owing to their
homogeneous geometrical configuration, we may take constant values for these
energies, $\varepsilon _{i}=\varepsilon _{\text{F}}$ and $t_{ij}=t$. Then,
the diagonal term yields just a constant, $\varepsilon _{\text{F}}N_{\text{C}%
}$, and can be neglected in the Hamiltonian (\ref{HamilTB}), where $N_{\text{%
C}}$ is the number of carbon atoms in a nanoribbon or nanodisk. This is
because there exists one electron per one carbon: The band-filling factor is
1/2.

We diagonalize the Hamiltonian (\ref{HamilTB}) explicitly to derive the
density of states for finite-length nanoribbons. It can be shown that the
determinant associated with the Hamiltonian (\ref{HamilTB})\ has a factor
such that%
\begin{equation}
\det \left[ \varepsilon I-H\left( N_{\text{C}}\right) \right] \propto
(\varepsilon -t)^{a\left( W,L\right) }(\varepsilon +t)^{a\left( W,L\right) },
\end{equation}%
implying the $a\left( W,L\right) $-fold degeneracy of the states with the
energy $\varepsilon =\pm t$, where 
\begin{subequations}
\begin{eqnarray}
a\left( 1,L\right) &=&2,1,2,1,2,1,2,1,2,1,\cdots , \\
a\left( 2,L\right) &=&1,1,0,2,0,1,1,0,2,0,\cdots , \\
a\left( 3,L\right) &=&2,0,2,0,2,0,2,0,2,0,\cdots , \\
a\left( 4,L\right) &=&1,2,0,3,0,2,1,1,2,0,\cdots .
\end{eqnarray}%
We have displayed the full spectrum for some examples in Fig.\ref{FDOS}.

One of our mains results is that there are no zero-energy states in
finite-length nanoribbons, though infinite-length nanoribbons have the flat
band consisting of degenerated zero-energy states\cite{Fujita,EzawaPRB}.

There are two interesting features in the energy spectra. First, the level
spacings are almost equal near the Fermi energy $\left\vert \varepsilon
\right\vert <t$, as shown in Fig.\ref{FDOS}. Second, the band gap decreases
inversely to the length, and zero-energy states emerge as $L\rightarrow
\infty $, as shown in Fig.\ref{FGap}. This is consistent with the fact that
infinite-length nanoribbons have the flat band made of degenerated
zero-energy states\cite{Fujita,EzawaPRB}. Hence, a sufficiently long
nanoribbon can be regarded practically as a metal. In this energy region the
energy spectrum is that of Dirac electrons\cite{GrapheneEx}. Hence we expect
to ascribe these features to the property of Dirac electrons, though
detailed mechanisms are yet to be studied in future works.

\begin{figure}[t]
\begin{center}
\includegraphics[width=0.8\textwidth]{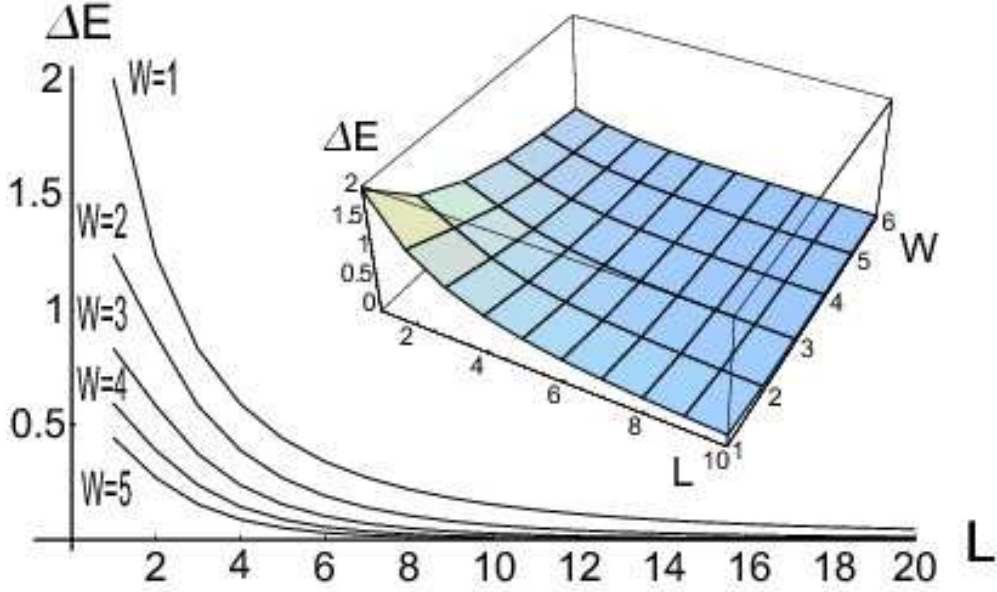}
\end{center}
\caption{(Color online) Band gap of zigzag nanoribbons with length $L$. The
horizontal axes is the length $L$ and the vertical axes is the energy gap $%
\Delta E$ in units of $t=3$eV. Inset: 3D plot of band gap as a function of $L
$ and $W$.}
\label{FGap}
\end{figure}

\textit{Graphene nanodisks}: We proceed to investigate the electronic
properties of a wide class of nanodisks. In particular, we are interested
whether there exist nanodisks which have zero-energy states. The basic
element of graphene nanodisks is a benzene. Every nanodisk can be
constructed by connecting several benzenes. There are a large variety of
graphene nanodisks, where typical examples are displayed in Fig.\ref%
{FigNanodisk}. We have studied hexagonal, parallelogrammic and trigonal
nanodisks with zigzag or armchair edges.

Diagonalizing the Hamiltonian (\ref{HamilTB}) explicitly, we have explicitly
constructed the energy spectrum for each of nanodisks. We have displayed the
density of states for several nanodisks with trigonal zigzag shape in Fig.%
\ref{FigNanoStair}(a) and parallelogrammic\ zigzag shape in Fig\ref%
{FigNanoStair}(b). We have checked that there exist zero-energy states only
in trigonal zigzag nanodisks. The emergence of zero-energy states in
graphene nanodisks is vary rare.

\begin{figure}[t]
\begin{center}
\includegraphics[width=0.6\textwidth]{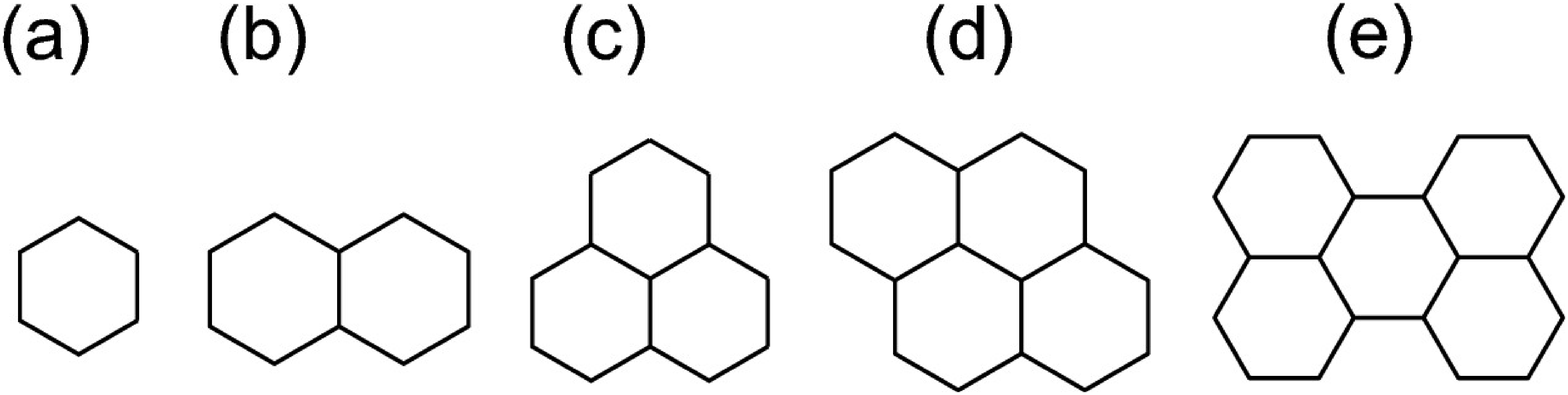}
\end{center}
\caption{Geometric configurations of typical graphene nanodisks. (a)
Benzene. (b) Naphthalene. (c) Trigonal zigzag nanodisk (phenalene). (d)
Pyrene. (e) Perylene.}
\label{FigNanodisk}
\end{figure}

\begin{figure}[t]
\begin{center}
\includegraphics[width=0.7\textwidth]{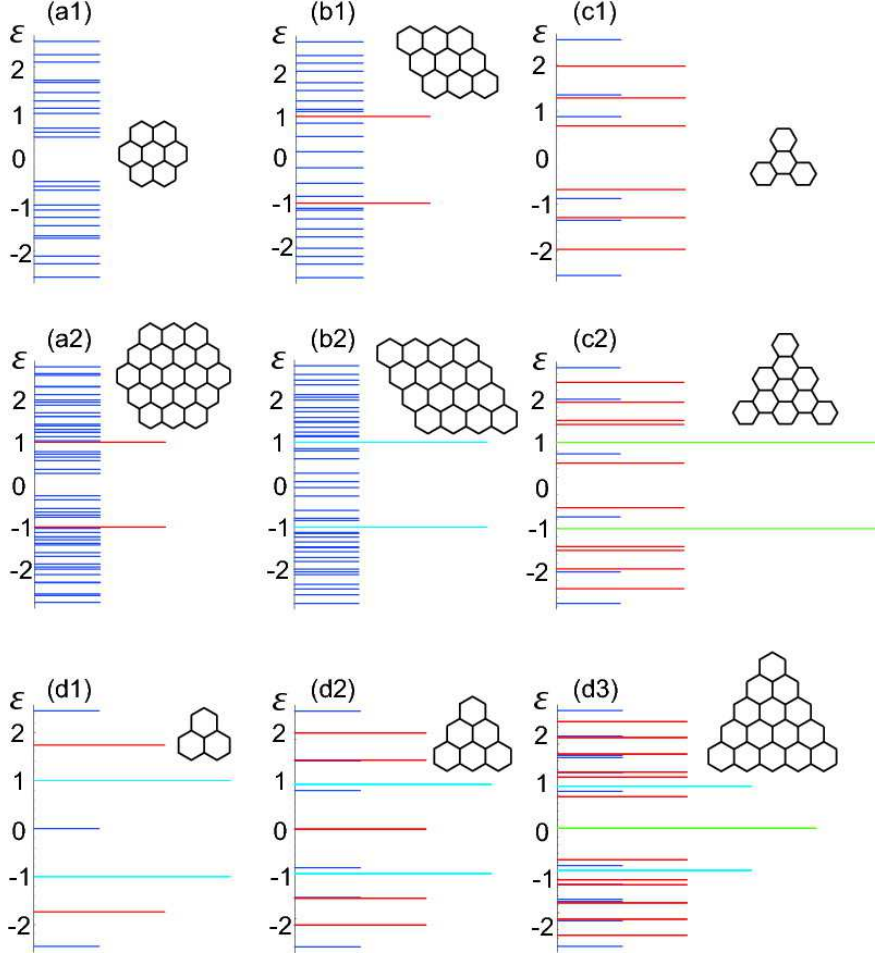}
\end{center}
\caption{(Color online) Density of states for typical zigzag nanodisks, The
horizontal axis is dgeneracy and the vertical axis is the energy $\protect%
\varepsilon $ in units of $t=3$eV. (a) Hexagonal zigzag nanodisks. (b)
Parallelogrammic\ zigzag nanodisks. (c) Trigonal armchair nanodisks. (d)
Trigonal zigzag nanodisks. There are degenerated zero-energy states in all
trigonal nanodisks, and they are metallic. There are no zero-energy states
in all other nanodisks, and they are semiconducting. }
\label{FigNanoStair}
\end{figure}

Trigonal zigzag nanodisks are prominent in their electronic properties
because there exist zero-energy states. We have found that the determinant
associated with the Hamiltonian (\ref{HamilTB})\ has a factor such that 
\end{subequations}
\begin{equation}
\det \left[ \varepsilon I-H\left( N_{\text{C}}\right) \right] \propto
\varepsilon ^{N}(\varepsilon -t)^{a\left( N\right) }(\varepsilon
+t)^{a\left( N\right) },
\end{equation}%
implying the $N$-fold degeneracy of the zero-energy states and the $a\left(
N\right) $-fold degeneracy of the states with the energy $\varepsilon =\pm t$%
, where%
\begin{equation}
a\left( N\right) =3,3,3,3,3,5,3,5,3,7,3,7,3,\cdots ,
\end{equation}%
for $N=1,2,3,\cdots $. Here, $N+1$ is the number of benzenes in one of the
edge of the trigonal nanodisk, which is related to the number of carbons by $%
N_{\text{C}}=N^{2}+6N+6$. It is remarkable that we can engineer nanodisks
equipped with an arbitrary number of degenerate zero-energy states.

\textit{Discussions}: Graphene nanodisks may be regarded as quantum dots
made by graphene. Finite-length graphene nanoribbons and graphene nanodisks
would be basic components of graphene nanocircuits.

I am very much grateful to Professors Y. Hirayama and K. Hashimoto for many
fruitful discussions on the subject.

\end{document}